%% file: SMP-15-008_temp.tex
\pdfoutput=1

\documentclass[11pt,twoside,a4paper,cmspaper,final,collab]{cms-tdr}
\def\svnVersion{429536P}\def\cmsCernNoTag{CERN-EP-2017-039}\def\cmsCernDate{\today}\def\cmsMessage{Published in the Journal of High Energy Physics as \href{http://dx.doi.org/10.1007/JHEP10(2017)072}{\doi{10.1007/JHEP10(2017)072.}}}

\begin{document}\cmsNoteHeader{SMP-15-008}

\hyphenation{had-ron-i-za-tion}
\hyphenation{cal-or-i-me-ter}
\hyphenation{de-vices}
\RCS$Revision: 425054 $
\RCS$HeadURL: svn+ssh://svn.cern.ch/reps/tdr2/papers/SMP-15-008/trunk/SMP-15-008.tex $
\RCS$Id: SMP-15-008.tex 425054 2017-09-11 12:21:01Z abelloni $
\newlength\cmsFigWidth
\ifthenelse{\boolean{cms@external}}{\setlength\cmsFigWidth{0.85\columnwidth}}{\setlength\cmsFigWidth{0.4\textwidth}}

\newcommand{\NA}{\ensuremath{\text{---}}\xspace}
\providecommand{\MGvNLO}{\MADGRAPH{}5\_a\MCATNLO\xspace}
\newcommand{\mt}{\ensuremath{m_\mathrm{T}}\xspace}
\newcommand{\fb}{\ensuremath{\,\text{fb}}}

\newcommand{\W}{\ensuremath{\cmsSymbolFace{W}}\xspace}
\newcommand{\V}{\ensuremath{\cmsSymbolFace{V}}\xspace}
\newcommand{\Wgg}{\ensuremath{\W\gamma\gamma}\xspace}
\newcommand{\ppWgg}{\ensuremath{{\Pp\Pp\to\W\gamma\gamma}}\xspace}
\newcommand{\Zgg}{\ensuremath{\Z\gamma\gamma}\xspace}
\newcommand{\ppZgg}{\ensuremath{\Pp\Pp\to\Z\gamma\gamma}\xspace}
\newcommand{\Meg}{\ensuremath{m_{\Pe\gamma}}}
\newcommand{\Megg}{\ensuremath{m_{\Pe\gamma\gamma}}}
\newcommand{\intlumi}{\ensuremath{19.4\fbinv}\xspace}
\newcommand{\fm}[1]{\ensuremath{f_{\mathrm{M},#1}}}
\newcommand{\ft}[1]{\ensuremath{f_{\mathrm{T},#1}}}
\newcommand{\chiso}{\ensuremath{I_{\text{ch}}}\xspace}
\newcommand{\eps}[2]{\epsilon_{\text{#1}}^{\text{#2}}}
\newcommand{\epsital}[2]{\epsilon_{#1}^{#2}}
\newcommand{\ftResult}{\ensuremath{-33.5 < \ft{0}/\Lambda^4 < 34.0\TeV^{-4}}}
\newcommand{\wggsignosig}{\ensuremath{2.6}\xspace}
\newcommand{\zggsignosig}{\ensuremath{5.9}\xspace}
\newcommand{\wggxsel}{\ensuremath{4.2\pm 2.0\stat\pm 1.6\syst\pm 0.1\lum\fb}\xspace}
\newcommand{\wggxsmu}{\ensuremath{6.0\pm 1.8\stat\pm 2.3\syst\pm 0.2\lum\fb}\xspace}
\newcommand{\wggxscomb}{\ensuremath{4.9\pm 1.4\stat\pm 1.6\syst\pm 0.1\lum\fb}\xspace}
\newcommand{\zggxsel}{\ensuremath{12.5\pm 2.1\stat\pm 2.1\syst\pm 0.3\lum\fb}\xspace}
\newcommand{\zggxsmu}{\ensuremath{12.8\pm 1.8\stat\pm 1.7\syst\pm 0.3\lum\fb}\xspace}
\newcommand{\zggxscomb}{\ensuremath{12.7\pm 1.4\stat\pm 1.8\syst\pm 0.3\lum\fb}\xspace}
\newcommand{\wggxscombsum}{\ensuremath{4.9\pm 2.1\fb}\xspace}
\newcommand{\zggxscombsum}{\ensuremath{12.7\pm 2.3\fb}\xspace}
\newcommand{\NNPDF}{{NNPDF-NLO}\xspace}
\newcommand{\CTTEN}{{CT10-NLO}\xspace}
\newcommand{\MSTW}{{MSTW-NLO}\xspace}
\newcommand{\CTEQSIX}{{CTEQ6L1}\xspace}

\cmsNoteHeader{SMP-15-008}

\title{Measurements of the \ppWgg and \ppZgg cross sections and limits on anomalous quartic gauge couplings at $\sqrt{s} = 8\TeV$}

\date{\today}

\abstract{
    Measurements are presented of \Wgg and \Zgg production in
    proton-proton collisions. Fiducial cross sections are reported
    based on a data sample corresponding to an integrated luminosity
    of \intlumi collected with the CMS detector at a center-of-mass
    energy of $8\TeV$. Signal is identified through the
    $\W \to \ell\nu$ and $\Z\to\ell\ell$ decay modes,
    where $\ell$ is a muon or an electron. The production of \Wgg
    and \Zgg, measured with significances of 2.6 and 5.9 standard
    deviations, respectively, is consistent with standard model
    predictions. In addition, limits on anomalous quartic gauge
    couplings in \Wgg production are determined in the context of a
    dimension-8 effective field theory. }

\hypersetup{%
pdfauthor={CMS Collaboration},%
pdftitle={Measurements of the pp to W gamma gamma and pp to Z gamma gamma cross sections
and limits on anomalous quartic gauge couplings at sqrt(s) = 8 TeV},%
pdfsubject={CMS},%
pdfkeywords={CMS, physics, triboson, aQGC}}

\maketitle
\section{Introduction}

Production of three-boson final states in proton-proton collisions is
predicted by the SU(2)$\times$U(1) gauge structure of the standard
model (SM). Cross sections for these processes include contributions
from quartic gauge couplings (QGCs), which are sensitive to new
phenomena that modify those couplings. In this paper, we present cross
section measurements for the \ppWgg and \ppZgg processes and a search
for anomalous QGCs (aQGCs). The $\W \to \ell \nu$ and $\Z\to\ell\ell$
decay modes are selected for analysis, where $\ell$ is a muon or an
electron. The cross sections are measured in fiducial regions that
are defined by selection criteria similar to those used to select
signal events. In particular, to avoid infrared divergences, minimum
photon transverse momenta \pt of 25 and 15\GeV are required in
the \Wgg and \Zgg measurements, respectively.
A dimension-8 effective field theory is used to
model aQGCs, which would enhance \Wgg production at high momentum
scales. The \Wgg and \Zgg processes were recently observed by the
ATLAS Collaboration~\cite{Aad:2015uqa,Aad:2016sau}
using 20.3\fbinv of integrated luminosity at $\sqrt{s}=8\TeV$. Cross
sections for \Wgg and \Zgg production have also been computed with QCD
corrections up to next-to-leading order (NLO) in
Ref.~\cite{WggNLOBozzi,ZggNLOBozzi}.

\section{The CMS detector and particle reconstruction}

The data used in these measurements amount to \intlumi collected in
2012 with the CMS detector at the CERN LHC in proton-proton collisions
at a center-of-mass energy of 8\TeV. A detailed description of the CMS
detector, together with definitions of the coordinate system and
relevant kinematic variables, can be found in
Ref.~\cite{Chatrchyan:2008zzk}. The central feature of the CMS
apparatus is a superconducting solenoid of 6\unit{m} internal
diameter, providing a magnetic field of 3.8\unit{T}. Within the field
volume are a silicon pixel and strip tracker, a lead tungstate crystal
electromagnetic calorimeter (ECAL), and a brass and plastic
scintillator hadron calorimeter (HCAL), each composed of a barrel and
two endcap sections. Extensive forward calorimetry utilizing a steel
absorber with embedded quartz fibers complements the coverage provided
by the barrel and endcap detectors. Muons are measured in
gas-ionization detectors embedded in the steel flux-return yoke
outside the solenoid.

The particle-flow (PF) algorithm~\cite{Sirunyan:2017ulk} reconstructs
and identifies five types of particles with an optimized combination
of information from the various elements of the CMS detector. Particle
flow candidates provide the basis for the selection and measurement of
muons, electrons, photons, jets, and the transverse momentum
imbalance. In addition, the isolation characteristics of identified
leptons and photons are measured using the \pt of PF charged hadrons,
neutral hadrons, and photons.

Muons are identified as tracks in the muon spectrometer that are
matched to tracks in the inner detector. Quality requirements are
placed on tracks measured in the inner detector and muon spectrometer,
as well as on the matching between them. Muons must also be isolated
from nearby PF candidates. Selected muons in the momentum range $20
<\pt < 100\GeV$ have a relative \pt resolution of 1.3--2.0\% in the
barrel ($\abs{\eta}<1.2$) and less than 6\% in the endcaps
($1.2<\abs{\eta}<2.4$)~\cite{Chatrchyan:2012xi}.

Photons and electrons are identified as clusters of energy deposits in
the ECAL. The energy of photons is directly obtained from the ECAL
measurement. Electrons are further identified by matching the ECAL
cluster to a track reconstructed in the inner detector. The momenta of
electrons are determined from a combination of the track momentum at
the primary interaction vertex, the energy of the corresponding ECAL
cluster, and the energy sum of all bremsstrahlung photons spatially
compatible with originating from the electron track. To take into
account electron bremsstrahlung in the inner-detector material, a
Gaussian sum filter algorithm~\cite{0954-3899-31-9-N01} is used to
measure the track momentum. The momentum resolution for electrons from
$\Z\to\Pep\Pem$ decays ranges from 1.7\% for electrons in the barrel
region to 4.5\% for electrons that begin to shower before the
calorimeter in the endcaps~\cite{Khachatryan:2015hwa}.

Electrons are selected in the \Wgg analysis using a multivariate
classifier based on the spatial distribution of the electron shower,
the energy deposited in the HCAL region matched to the ECAL shower,
and the quality of the inner-detector track. Electrons are selected
in the \Zgg analysis by imposing looser requirements on the same
variables, yielding improved signal acceptance.
In both cases, electrons passing the selection must also be isolated
from nearby PF candidates.

Photons are identified using a selection that requires a narrow shower
in the ECAL, minimal energy deposited in the HCAL region matched to
the ECAL shower, and isolation from nearby PF candidates. Separate
isolation requirements are placed on the energies of PF charged
hadrons, neutral hadrons, and photons. Photons that convert to an
electron-positron pair are included and the same selection criteria
are applied. The energy resolution is about 1\% in the barrel section
of the ECAL for unconverted or late converting photons in the tens
of \GeVns energy range. The remaining barrel photons have a resolution
of about 1.3\% up to a pseudorapidity of $\abs{\eta} = 1$, rising to
about 2.5\% at $\abs{\eta} = 1.4$. In the endcaps, which cover a
pseudorapidity of $1.5 < \abs{\eta} < 2.5$, the resolution of
unconverted photons is about 2.5\%, while converted photons have a
resolution between 3 and 4\%~\cite{CMS:EGM-14-001}.

The transverse momentum imbalance vector \ptvecmiss is defined as the
projection on the plane perpendicular to the beams of the negative
vector sum of the \ptvec of all reconstructed PF candidates in the
event. Its magnitude is referred to as \ptmiss. Corrections to the
energy scale and resolution of jets, described
in~\cite{Khachatryan:2016kdb}, are propagated to the calculation
of \ptmiss.

\section{Event selection}
Events are recorded using single-lepton triggers for the \Wgg
selection and dilepton triggers for the \Zgg
selection~\cite{Khachatryan:2016bia}. The single-lepton triggers
have \pt thresholds of 24 and 27\GeV for muons and electrons,
respectively. The dimuon and dielectron triggers both have \pt
thresholds of 17 and 8\GeV on the leading and subleading leptons,
respectively. To ensure uniform trigger efficiency, reconstructed
leptons are required to have \pt above the trigger thresholds. The \pt
requirement is determined by measuring the efficiency of the trigger
as a function of \pt and selecting the value at which the efficiency
becomes approximately independent of \pt. For the \Wgg (\Zgg) analysis
the muons and electrons must have minimum \pt of 25\,(10) and
30\,(20)\GeV, respectively.

Events selected for the \Wgg analysis must have one muon or electron
and two photons. Each photon is required to have
\pt greater than 25\GeV. Events are removed if a second lepton
is present having \pt above 10\GeV. All reconstructed leptons and
photons must be separated from each other by $\Delta R > 0.4$, where
$\Delta R = \sqrt{\smash[b]{\left(\Delta\phi\right)^{2}
+ \left(\Delta\eta\right)^{2} }}$ and $\phi$ is the azimuthal angle.
To identify leptonic $\W$ boson decays and remove backgrounds not
having genuine \ptmiss, the transverse mass, defined as
\begin{equation*}
\mt = \sqrt{
    2p^{\ell}_\mathrm{T}\ptmiss (1 - \cos[\phi(\vec{p}^{\ell}_\mathrm{T})
- \phi(\ptvecmiss)])},
\end{equation*}

is required to be greater than 40\GeV; $p^{\ell}_\mathrm{T}$ denotes
the \pt of the lepton. In the electron channel, additional criteria
are imposed to reject background events arising from \Z boson decays
to electrons in which only one electron is correctly identified, the
other is misidentified as a photon, and an additional prompt photon is
present in the event. Both photons are required to pass an electron
veto that rejects photons that match to tracks in the pixel
detector. This requirement decreases the signal efficiency by removing
converted photons, which are commonly matched to tracks in the pixel
detector. However, the background contamination from electrons is
further decreased by a factor of two. Events are also removed if the
invariant mass of any combination of the electron and one or both
photons is near the \Z boson mass. In particular, events are removed
if they have $86 < \Meg < 96\GeV$ for either combination of a photon
with the electron, or if $86 < \Megg < 96\GeV$, in which case one
photon is likely to be from final-state radiation (FSR).

Events selected for the \Zgg analysis must have two electrons or muons
of opposite charge and two photons. Each photon is required to have a
minimum \pt of 15\GeV. Photons are required to pass an electron veto
that has a higher signal efficiency than that used in the electron
channel of the \Wgg analysis. All reconstructed leptons and photons
must be separated from each other by $\Delta R > 0.4$. The dilepton
invariant mass must be greater than 40\GeV to remove backgrounds that
have low dilepton invariant masses.

In both analyses, photons reconstructed in the barrel and endcaps are
treated separately. The geometry of the ECAL differs between the
barrel and endcaps and therefore different selection criteria are
imposed for each case. Photons that are reconstructed in the endcaps
are more likely to originate from misidentified jets. Events in which
both reconstructed photons are in the endcaps are not considered in
the analysis because of the unfavorable signal-to-background ratio.

\section{Signal and background simulation\label{sec:mcsamples}}

Simulated events are generated at NLO for the \Wgg and \Zgg
signals. These samples are generated with
\MGvNLO(v5 2.2.2)~\cite{Alwall:2014hca} using the
\NNPDF(v.3.0)~\cite{Ball:2014uwa} parton distribution functions (PDFs),
and showered with \PYTHIA(v.8.1)~\cite{Sjostrand:2007gs} using the
Monash tune~\cite{Skands:2014pea}.

Events are generated that model the aQGC signals and the diboson and
triboson backgrounds at leading order (LO) using \MADGRAPH(v5 2.2.2)
using the \CTEQSIX~\cite{CTEQ6PDF} PDF set, and then showered
with \PYTHIA (v.6.4)~\cite{Sjostrand:2006za} Z2*
tune~\cite{Chatrchyan:2013gfi}.

Simulated aQGC events are assigned a set of weights, each of which
reproduces the effect of an anomalous QGC. The weights are obtained by
loading models of effective theories, provided in the Universal
FeynRules Output format~\cite{Degrande:2011ua}, into the event
generator. The diboson and triboson predictions are normalized to the
NLO cross section predictions obtained
with \MCFM(v.6.6)~\cite{Campbell:2010ff} and \MGvNLO(v5 2.2.2),
respectively. All $\tau$ leptons included in samples showered
with \PYTHIA are decayed
with \TAUOLA(v.1.1.1a)~\cite{Davidson:2010rw}.

The influence of additional proton-proton collisions in data events
(pileup) is corrected by adding minimum-bias collisions to the
simulated events. The number of added pileup collisions follows a
distribution that is similar to the distribution observed in data and
an additional weight is applied such that the simulated pileup
distribution accurately represents the data. Finally, all simulated
samples are passed through a detailed \GEANTfour
simulation~\cite{Agostinelli:2002hh} of the CMS detector.

Corrections for differences between the simulation and the data in the
selection efficiencies of muons, electrons, and photons and in the
trigger efficiencies are determined using the tag-and-probe method and
applied to the simulated events. Differences in the momentum scale of
muons, electrons, and photons are determined from the \Z boson line
shape, and the simulation is corrected to agree with the data.

\section{Background estimation}

The main background contribution in both analyses consists of events
in which one or two jets are misidentified as photons. In fact, while
the photon shower and isolation requirements are designed to reject
misidentified jets, the relatively large production rate of
electroweak bosons with jets leads to a large contribution of jets
misidentified as photons. A jet is commonly misidentified as a photon
when it contains a neutral meson that decays to
overlapping photons. If the photons carry a large fraction of the jet
energy such that the other hadronization products have low momentum,
the reconstructed photon can pass the isolation requirements.
The probability for a jet to be misidentified as a photon is sensitive
to how jets interact with the detector and is therefore difficult to
predict with simulation. Moreover, the generation of a sufficiently
large simulated sample is impractical because of the large rejection
factor obtained through the photon identification criteria. A
data-based method is therefore used to estimate the contamination from
this source.

The background estimate is based on an analysis of the two-dimensional
distribution of the charged hadron isolation variables
$I_{\text{ch},1}$ and $I_{\text{ch},2}$ of the leading and subleading
photon candidates, respectively. The isolation \chiso is defined as
the scalar \pt sum of charged hadron PF candidates having $\Delta R <
0.3$ with respect to the photon candidate. Charged hadron PF
candidates are required to have energy deposits in the HCAL and
originate from the primary vertex, defined as the vertex with the
highest sum of squared transverse momenta of its associated
tracks~\cite{Chatrchyan:2014fea}. Prompt photons have low values
of \chiso while jets that are misidentified as photons tend to have
larger values. The distribution of $I_{\text{ch},1}$ versus
$I_{\text{ch},2}$ (a ``template'') is determined for each of the four
sources of diphoton candidates: prompt-prompt (PP), prompt-jet (PJ),
jet-prompt (JP), and jet-jet (JJ). The PP template represents the
signal, while the PJ and JP templates represent background events
having one prompt photon, and the JJ template represents background
events having no prompt photons. Each template consists of four
bins. The distribution of \chiso\ is divided into a ``tight'' region
and a ``loose'' control region for each of the two photons. The tight
region contains photon candidates that satisfy the nominal \chiso\
criterion, while the loose region contains photon candidates that fail
the nominal, but pass a less stringent requirement. The value of the
less stringent requirement is chosen such that candidates in the loose
region are enriched in photon-like jets that are independent of, but
sufficiently similar to those that contaminate the signal region. The
four-bin structure of the templates provides discrimination between
prompt photons and jets and allows for a straightforward matrix
equation solution, taking account of correlations between
$I_{\text{ch},1}$ and $I_{\text{ch},2}$. The contribution of each
source is determined from control data samples. Three control data
samples are formed from the combinations of the tight and loose
regions: tight-loose (TL) and loose-tight (LT), where one photon
passes the requirement and the other fails, and loose-loose (LL),
where both photons fail the requirement. The signal region is labeled
tight-tight (TT). The TL and LT regions are treated separately to take
into account differences in photon \pt and differences between photons
that are reconstructed in the barrel and endcaps. The normalizations
of the four sources of photon candidates are determined through the
matrix equation

\begin{equation}
\left(
\begin{array}{c}
 N_\text{TT} \\[1.5mm]
 N_\text{TL} \\[1.5mm]
 N_\text{LT} \\[1.5mm]
 N_\text{LL}
\end{array}
\right)
 =
\left(
\begin{array}{cccc}
 \eps{PP}{TT} & \eps{PJ}{TT} & \eps{JP}{TT} & \eps{JJ}{TT}\\[1.5mm]
 \eps{PP}{TL} & \eps{PJ}{TL} & \eps{JP}{TL} & \eps{JJ}{TL}\\[1.5mm]
 \eps{PP}{LT} & \eps{PJ}{LT} & \eps{JP}{LT} & \eps{JJ}{LT}\\[1.5mm]
 \eps{PP}{LL} & \eps{PJ}{LL} & \eps{JP}{LL} & \eps{JJ}{LL}
\end{array}
\right)
\left(
\begin{array}{c}
 \alpha_\text{PP} \\[1.5mm]
 \alpha_\text{PJ} \\[1.5mm]
 \alpha_\text{JP} \\[1.5mm]
 \alpha_\text{JJ}
\end{array}
\right),
\label{eq:matrix}
\end{equation}
where $N_{XY}$ is the observed number of events in region $XY$,
$\epsital{AB}{XY}$ is the probability for an event from source $AB$ to
appear in region $XY$, as determined from the templates, and
$\alpha_{AB}$ is the normalization of source $AB$. Each column in the
matrix corresponds to the four bins from one template, and the entries
in the column sum to unity by construction. The predicted number of
events from source $AB$ reconstructed in region $XY$ is given by the
product $\alpha_{AB}\,\epsital{AB}{XY}$. The final background estimate is
the sum of the contributions from the sources involving at least one
jet:
\begin{equation*}
\alpha_\text{PJ}\,\eps{PJ}{TT} +
\alpha_\text{JP}\,\eps{JP}{TT} +
\alpha_\text{JJ}\,\eps{JJ}{TT}.
\end{equation*}

Templates are constructed from both Monte Carlo (MC) simulation and
data control samples. This procedure is applied separately for
different ranges of photon \pt and $\eta$. The templates for the PP,
PJ, and JP sources are determined from prompt and jet \chiso
distributions obtained from single-photon events. 
The single-photon \chiso distributions are binned in the same manner
as the templates to create two-bin distributions representing the
leading and subleading photon. Products of the two-bin distributions
corresponding to the leading and subleading photons are used to
determine the four-bin templates, the entries of which appear in
Eq.~(\ref{eq:matrix}).

The \chiso distribution for prompt photons is taken from simulated
$\W\gamma$ events. Simulated events are required to contain one
reconstructed photon that matches a photon in the generator record
within $\Delta R = 0.2$ and passes all selection criteria except
the \chiso requirement. The distributions obtained from simulation are
validated with data events in which an FSR photon is identified in
a \Z boson decay to $\mu^+\mu^-$. To ensure that the photon results
from FSR, the three-body invariant mass is required to be consistent
with the \Z boson mass and the photon must be within $\Delta R = 1$ of
a muon.  The available data sample is adequate to make this comparison
for photons with \pt up to 40\GeV, and good agreement is observed
between data and simulation.  An uncertainty of 10--20\% is applied,
depending on the photon \pt and $\eta$, to take into account the
observed differences and for the extrapolation to higher photon \pt.

The \chiso distribution for jets is taken from data. For this
purpose, events are selected that contain two reconstructed muons with
invariant mass consistent with the \Z boson mass and a reconstructed
photon that passes all selection criteria except the
\chiso requirement. To exclude genuine photons from FSR,
the photon is required to be separated from each muon by $\Delta R >
1$. The remaining contribution from prompt photons is subtracted using
the prediction from a sample of simulated $\Z\gamma$ events normalized
to its production cross section calculated at next-to-next-to-leading
order~\cite{Grazzini:2015nwa}. This normalization is checked with a
control data sample similar to that used to validate the \chiso\
distribution for prompt photons. Based on this comparison, a
systematic uncertainty of 20\%, dominated by the statistical
uncertainty in the control sample, is assessed to the $\Z\gamma$
normalization.

Events that have two jets misidentified as photons represent
approximately 30\% and 10\% of the total misidentified jet background
in the \Wgg and \Zgg analyses, respectively. In such events,
nonnegligible correlations exist between the leading and subleading
photons. These correlations originate from the event activity that
affects the measured isolation energies of both photons. The JJ
templates are therefore determined from a sample of candidate diphoton
events in data that is independent of the signal region. For this
selection, the requirement on the ECAL transverse shower shape is
inverted and the PF photon isolation requirement is relaxed. This
procedure can result in a bias through correlations between the ECAL
shower shape and the isolation.  The systematic uncertainties are
estimated by varying the maximum value of the relaxed requirements on
the PF photon isolation. The largest deviation is taken as an estimate
of the systematic uncertainty, which is approximately 10\%. 
Using this method, rather than treating the photons as uncorrelated, 
increases the contribution from jet-jet events, 
which increases the estimated background by as much as 30\%.

The total uncertainties in the estimated background contamination from
misidentified jets are 19\% and 28\% for the muon and electron $\Wgg$
channels, respectively, and 14\% for the muon and electron $\Zgg$
channels. These uncertainties take into account systematic effects in
the derivation of the probabilities for prompt photons and jets
described above, and statistical uncertainties in the observed data.
The larger uncertainty in the electron channel of the \Wgg analysis
results from the smaller amount of data as well as larger systematic
variations in the JJ template determination.

In the electron channel of the \Wgg analysis, a nonnegligible
contamination is present from $\Z({\to}\Pe\Pe)\gamma$ events in which
an electron is misidentified as a photon. An electron veto based on
pixel tracks is used as a discriminating variable to determine a
misidentification ratio. This ratio relates the number of events that
fail the electron veto to the number that pass. The misidentification
ratio is determined as a function of \pt and $\eta$ in a control
sample of data enriched in single \Z boson events that have one
reconstructed electron and one photon. The contamination in the signal
region is obtained by multiplying the observed number of events
outside the \Z boson mass window where one photon fails the electron
veto by the misidentification ratio.
The number of electrons resulting from \Z boson decays is extracted
from a fit to the $\Pe\gamma$ invariant mass distribution using a \Z
boson line shape determined from simulation and a background function
that models the contribution from events without a \Z boson. The
misidentification ratio is 0.01--0.03, depending on the \pt and $\eta$
of the photon. A systematic uncertainty of 10\% in the
misidentification ratio is determined from a closure test in
simulation. The contamination from misidentified jets in the control
samples is determined using the method described above and subtracted
from the data. This contamination is approximately 10\% for events in
which both photons are in the barrel and 20\% for the remaining
events.

Additional background contributions involving prompt photons are
determined using MC simulations. The simulated events are corrected
for observed differences in the selection efficiencies between data
and simulation of electrons, muons, and photons and in the trigger
efficiencies. In the \Wgg analysis, the contamination from \Zgg is
estimated using the \Zgg MC sample described in
Section~\ref{sec:mcsamples}. The \Zgg contamination constitutes about
90\% of the background that contains two prompt photons. The simulated
sample is normalized to the NLO cross section with an uncertainty of
12.5\%, based on the uncertainty in the theoretical prediction and
differences in identification and reconstruction efficiencies between
data and simulation. Contributions of less than an event per channel
from top quark production and other multiboson processes, including
$\ttbar\gamma\gamma$, $\cPqt\W\gamma\gamma$, and $\V\V\gamma\gamma$,
where $\V$ is a $\W$ or $\Z$ boson, are present in both the \Wgg
and \Zgg final states. These background sources are estimated using
leading-order MC simulation. A systematic uncertainty of 20\% is
applied to the sum of these contributions to take into account
higher-order corrections and differences in identification and
reconstruction efficiencies between data and simulation.

\begin{figure}[bph!]
\centering
\includegraphics[width=0.5\textwidth]{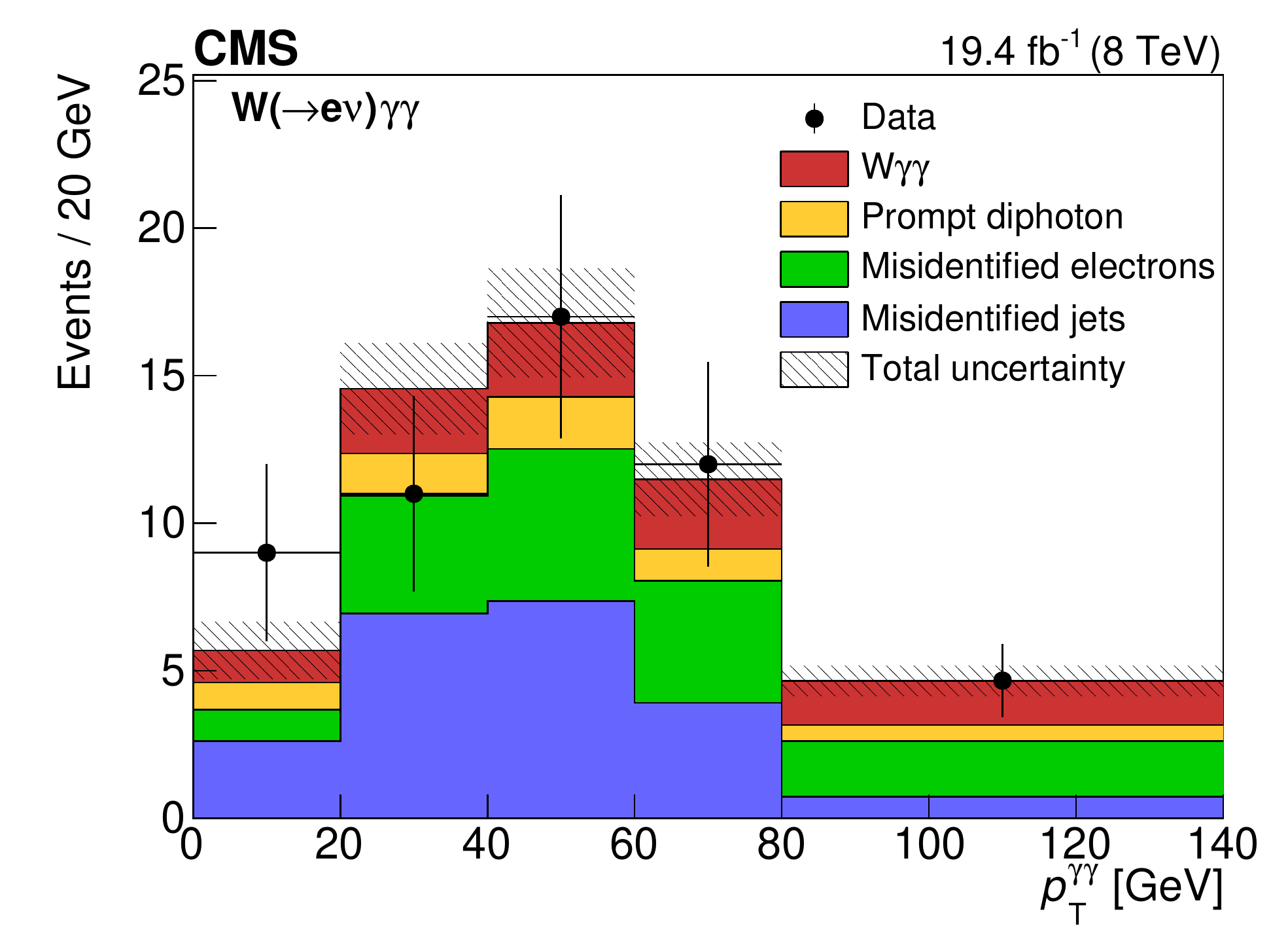}%
\includegraphics[width=0.5\textwidth]{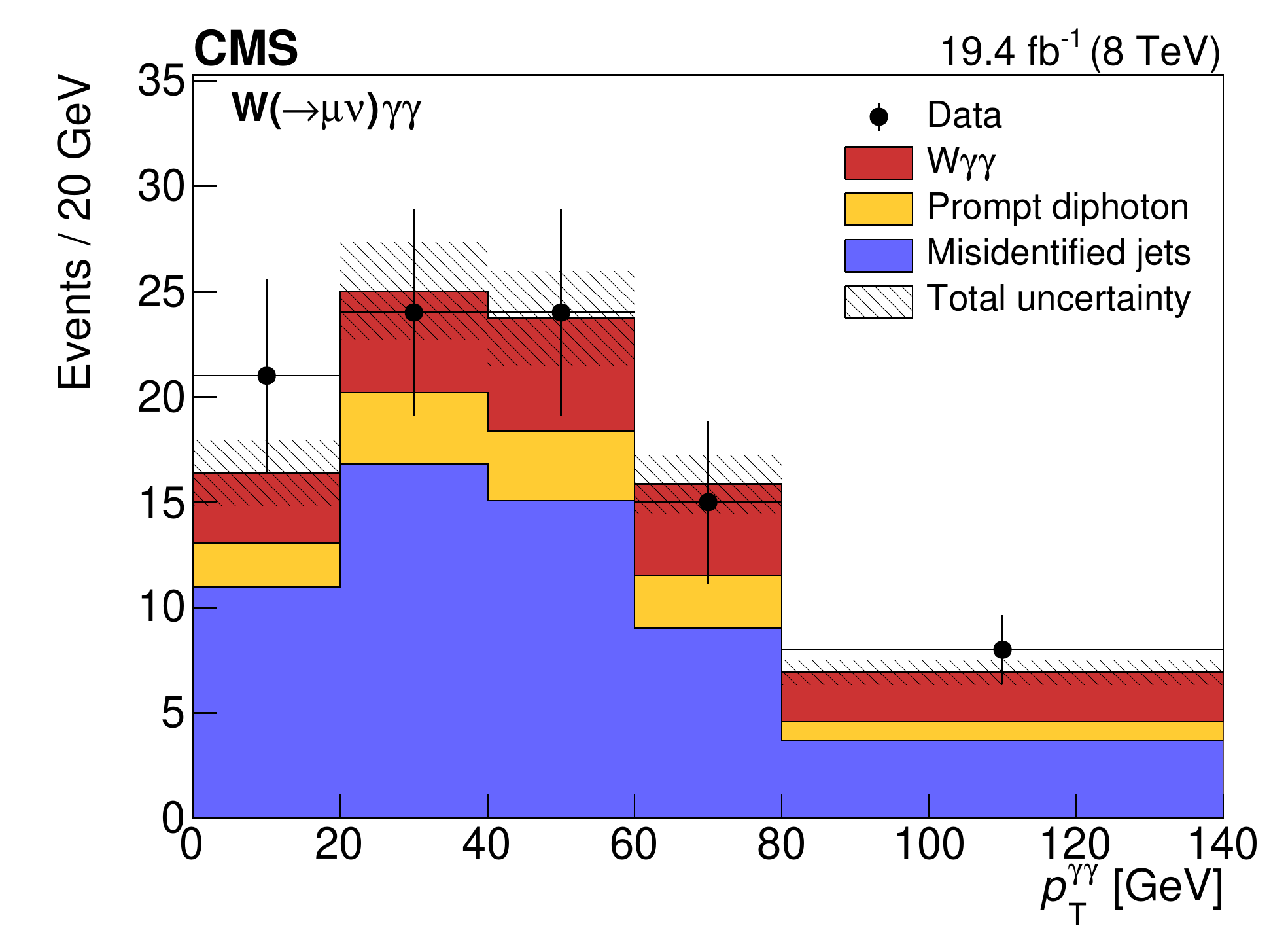}

\includegraphics[width=0.5\textwidth]{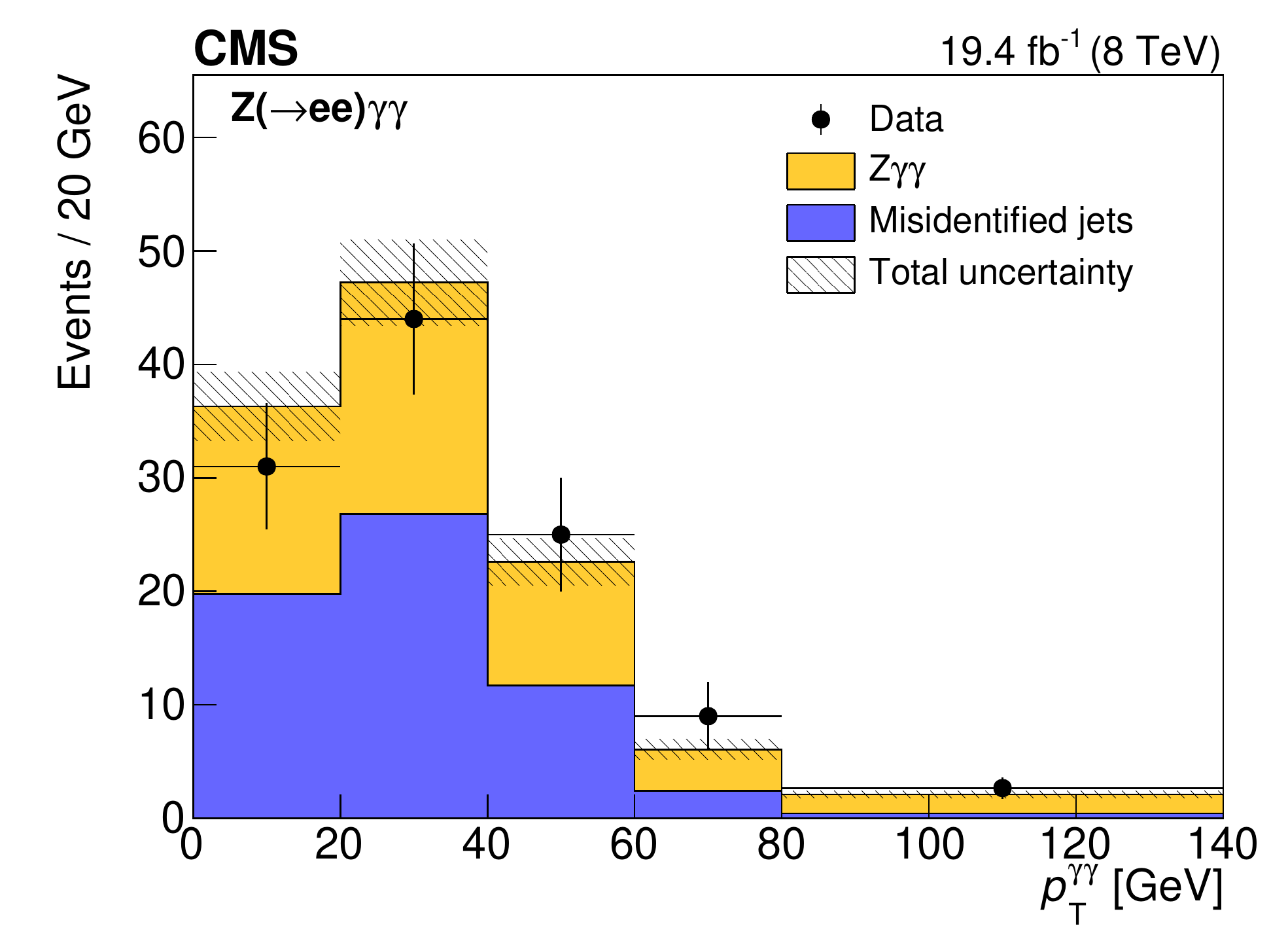}%
\includegraphics[width=0.5\textwidth]{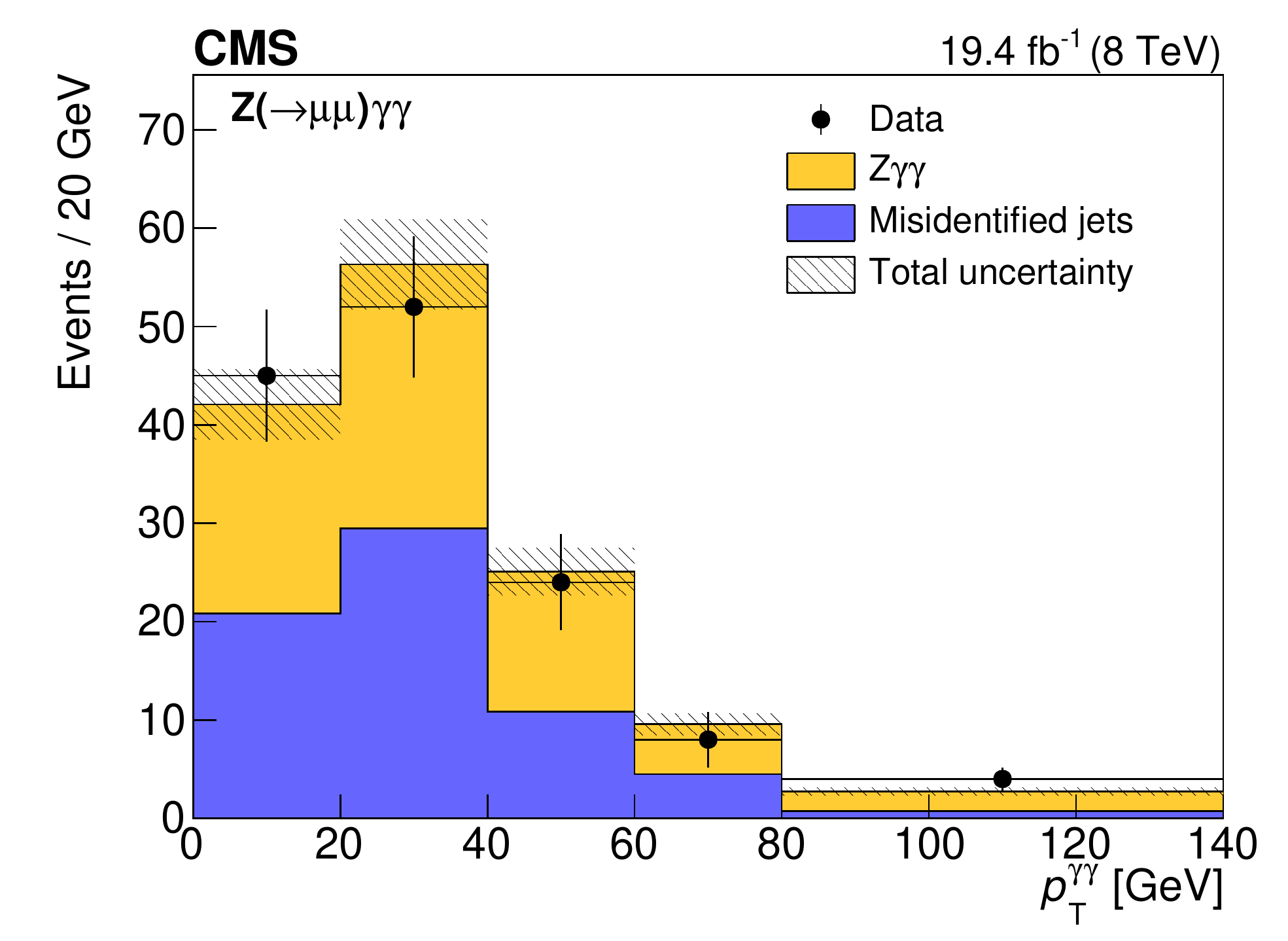}
\caption{\label{fig:wggsr}
  Distributions of the diphoton \pt for the \Wgg (upper) and \Zgg
  (lower) analyses, in the electron~(left) and muon~(right)
  channels. The points display the observed data and the histograms
  show the predictions for the background and signal. The indicated
  uncertainties in the data points are calculated using Poisson
  statistics. The hatched area displays the total uncertainty in the
  sum of these predictions. The predictions for electrons and jets
  misidentified as photons are obtained with data-based methods. The
  remaining background and signal predictions are derived from MC
  simulation. The last bin includes all events in which the
  diphoton \pt exceeds 80\GeV.}
\end{figure}

\begin{figure}[!ht]
\centering
\includegraphics[width=0.5\textwidth]{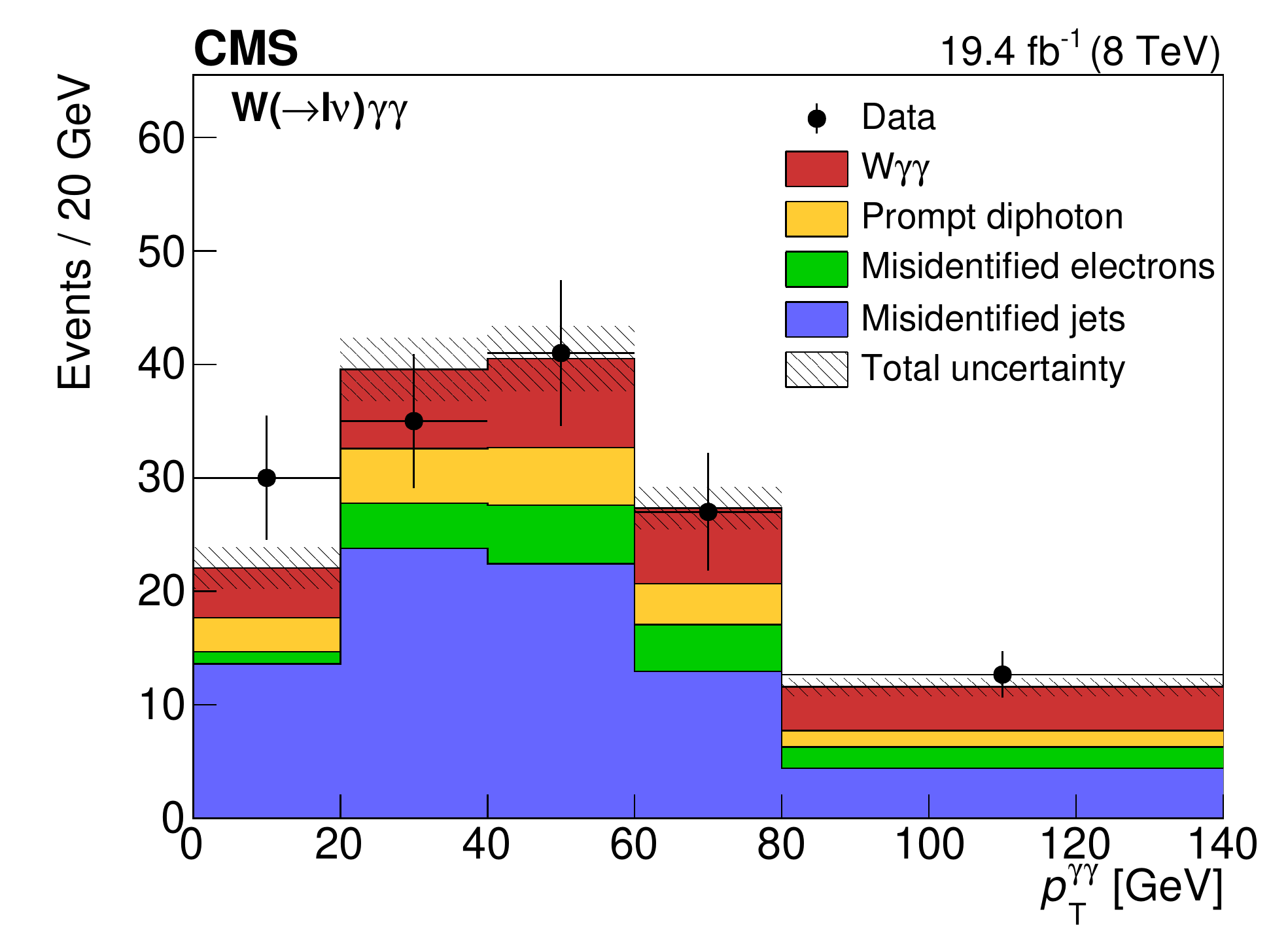}%
\includegraphics[width=0.5\textwidth]{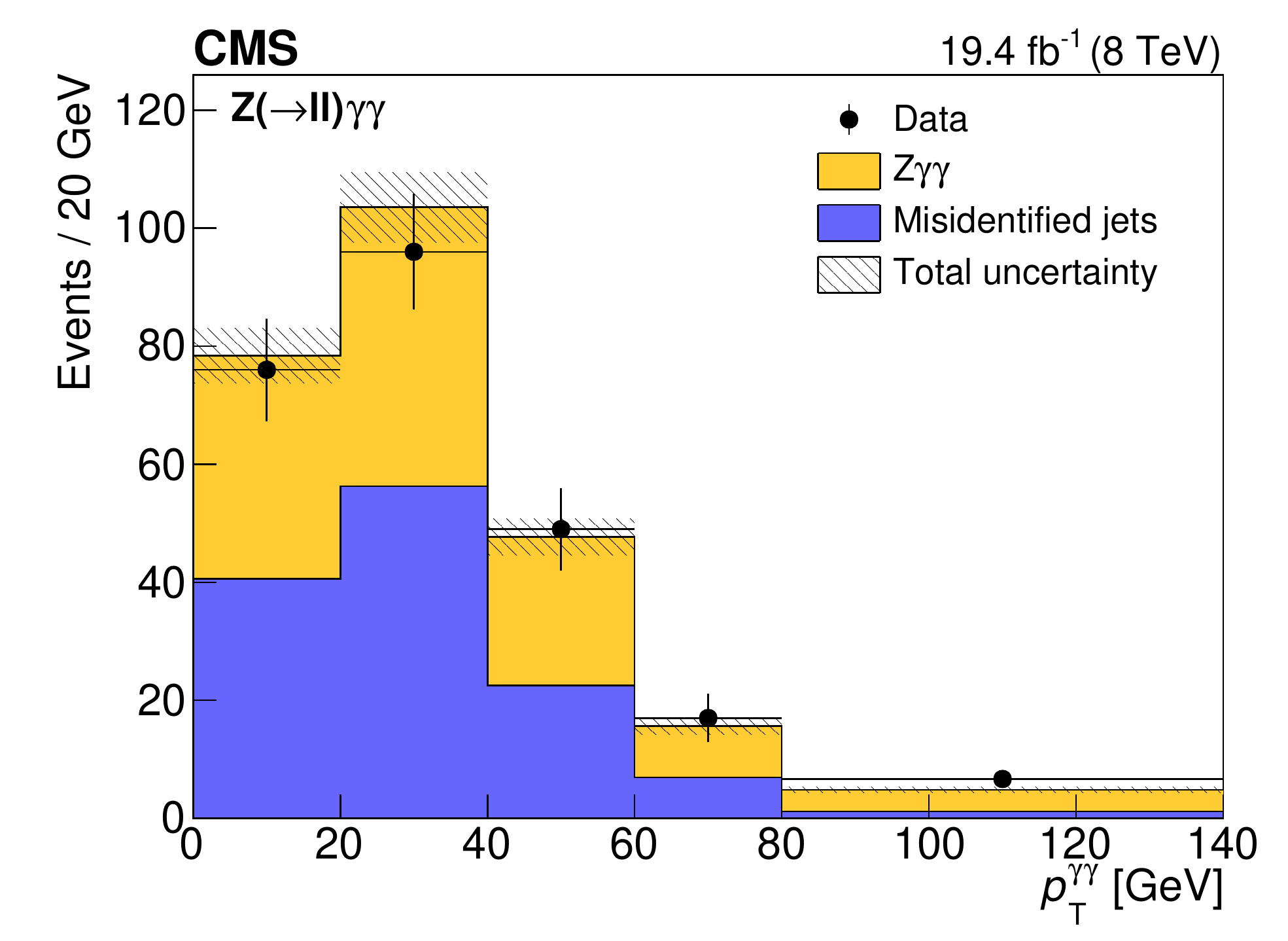}
\caption{\label{fig:wzggsr}
  Distributions of the diphoton \pt for the \Wgg~(left)
  and \Zgg~(right) analyses with the electron and muon channels
  summed. The points display the observed data and the histograms give
  the predictions for the background and signal. The indicated
  uncertainties in the data points are calculated using Poisson
  statistics. The hatched area displays the total uncertainty in the
  sum of these predictions. The predictions for electrons and jets
  misidentified as photons are obtained with data-based methods. The
  remaining background and signal predictions are derived from MC
  simulation. The last bin includes all events in which the
  diphoton \pt exceeds 80\GeV.}
\end{figure}

Table~\ref{tab:yields} summarizes the background predictions and the
observed numbers of events, which are consistent with the presence of
signal. Figure~\ref{fig:wggsr} shows the diphoton \pt distribution
with the predicted background, signal, and observed data for the \Wgg
and \Zgg analyses, separately in the electron and muon channels.
Figure~\ref{fig:wzggsr} shows the same distributions with the electron
and muon channels combined. The \Wgg and \Zgg signals are observed
with significances of \wggsignosig and \zggsignosig standard
deviations, respectively. The significances of the signals are
calculated using a profile likelihood that considers the observed data
and predicted backgrounds in each of the muon and electron
channels. In this calculation, separate categories are defined for
events having both photons in the barrel and only one photon in the
barrel, to take advantage of the higher signal-to-background ratio in
the first category as compared to the second.

\begin{table}[tb]
\begin{center}
\topcaption{\label{tab:yields}
  Background composition, expected signal, and observed yields in the \Wgg
  (upper) and \Zgg (lower) analyses.}
\newcolumntype{x}{D{,}{\,\pm\,}{3.3}}
\begin{tabular}{lxx}
\hline\hline
\Wgg & \multicolumn{1}{c}{Electron channel} & \multicolumn{1}{c}{Muon channel} \\
\hline
$\text{Jet}\to\gamma$ misidentification      & 22 , 6 & 63 , 12 \\
$\text{Electron}\to\gamma$ misidentification & 20 , 2 & \multicolumn{1}{c}{\NA}          \\
Prompt diphoton                       & 7 , 1  & 14 , 2 \\ \hline
Total background                      & 49 , 6 & 77 , 12 \\
\hline
Expected signal                       & 13 , 1 & 25 , 3  \\
\hline
Data                                  & \multicolumn{1}{c}{63}         & \multicolumn{1}{c}{108}         \\
\hline
\hline
\Zgg & \multicolumn{1}{c}{Electron channel} & \multicolumn{1}{c}{Muon channel}\\
\hline
$\text{Jet}\to\gamma$ misidentification & 62 , 8    & 68 , 9    \\
Prompt diphoton                  & 0.3 , 0.1 & 0.6 , 0.2 \\ \hline
Total background                 & 62 , 8    & 69 , 9    \\
\hline
Expected signal                  & 56 , 8    & 73 , 10   \\
\hline
Data                             & \multicolumn{1}{c}{117}           & \multicolumn{1}{c}{141}           \\
\hline
\hline
\end{tabular}
\end{center}
\end{table}

\section{Cross section measurements}

The \Wgg and \Zgg cross sections are measured within fiducial regions
identified by the selection criteria listed in
Table~\ref{tab:fid_def}. The acceptances of the fiducial regions for
the signal processes as well as their reconstruction and selection
efficiencies are determined using the signal MC samples described in
Section~\ref{sec:mcsamples}. In the MC simulation, photons are
required to satisfy a Frixione isolation requirement with a distance
parameter of 0.05~\cite{Frixione}. The fiducial selection criteria are
applied to the generated lepton four-momenta after a correction for
FSR, which is obtained by adding to the generated four-momentum of
each lepton the generated four-momenta of all photons within $\Delta R
< 0.1$. The fiducial cross sections are defined for \W and \Z boson
decays to a single lepton family ($\ell$).

Leptonic decays of $\tau$ leptons resulting from \W and \Z decays also
contribute to signal events. Based on simulation the $\tau$ lepton
contamination in the \Wgg fiducial region is approximately 2.5\%,
while in the \Zgg fiducial region it is less than 1\%. The combined
acceptances and efficiencies, after subtracting the $\tau$ lepton
contribution, are 17.3 and 26.7\% for the electron and muon channels
of the \Wgg analysis, respectively, and 22.5 and 29.1\% for the \Zgg
analysis.

Uncertainties in the acceptances result from uncertainties in the PDFs
of the proton, the perturbative QCD renormalization and factorization
scales, the number of additional pileup interactions, and the
selection efficiencies of leptons, photons, and \ptmiss. The PDF
uncertainties are evaluated by comparing the acceptances obtained with
the \NNPDF error sets and between the nominal \NNPDF set and the \MSTW
2008~\cite{Martin:2009iq} and \CTTEN~\cite{Lai:2010vv} PDF sets. The
maximum deviation from the nominal acceptance is taken as a systematic
uncertainty. The uncertainties related to the renormalization and
factorization scales are evaluated by varying them independently by
factors of 0.5 and 2. The largest variation is applied as a systematic
uncertainty. The uncertainty in the pileup distribution is evaluated
by varying the assumed minimum-bias cross section by
$\pm$5\%. Uncertainties in the selection efficiencies of electrons,
muons, and photons and in the trigger requirements are derived from
uncertainties in the tag-and-probe analyses. Estimates of the energy
scale uncertainty for the electron, photon, and muon are made from
comparisons of the Z boson line shape between data and simulation.
Uncertainties in the \ptmiss energy scale are estimated by propagating
the energy scale uncertainty for each object used in the \ptmiss
calculation. The total uncertainties in the combined acceptances and
efficiencies are 1--2\%. The integrated luminosity used for these
measurements is 19.4\fbinv with an uncertainty of
2.6\%~\cite{CMS-PAS-LUM-13-001}. A summary of the systematic
uncertainties affecting the \Wgg and \Zgg fiducial cross section
measurements is reported in Table~\ref{tab:wzfid_uncert}.

\begin{table}[tbph]
\begin{center}
\topcaption{\label{tab:fid_def}
  Fiducial region definitions for the \Wgg analysis (upper) and \Zgg analysis
  (lower). The transverse mass \mt is defined as in the event selection, but
  with \ptmiss replaced by the neutrino transverse momentum.}
\renewcommand*{\arraystretch}{1.2}
\begin{tabular}{c}
\hline \hline
{Definition of the \Wgg fiducial region} \\
\hline
$\pt^{\gamma} > 25\GeV$, $\abs{\eta^\gamma} < 2.5$\\
$\pt^\ell > 25\GeV$, $\abs{\eta^\ell} < 2.4$\\
One candidate lepton and two candidate photons \\
$\mt > 40\GeV$ \\
$\Delta R(\gamma, \gamma) > 0.4$ and $\Delta R(\gamma, \ell) > 0.4$\\
\hline \hline
{Definition of the \Zgg fiducial region} \\
\hline
$\pt^{\gamma} > 15\GeV$, $\abs{\eta^\gamma} < 2.5$\\
$\pt^\ell > 10\GeV$, $\abs{\eta^\ell} < 2.4$\\
Two oppositely charged candidate leptons and two candidate photons \\
leading $\pt^{\ell} > 20\GeV$\\
$m_{\ell\ell} > 40\GeV$ \\
$\Delta R(\gamma, \gamma) > 0.4$, $\Delta R(\gamma, \ell) > 0.4$, and $\Delta R(\ell, \ell) > 0.4$\\
\hline \hline
\end{tabular}
\end{center}
\end{table}

\begin{table}[tbph]
\begin{center}
\topcaption{\label{tab:wzfid_uncert}
  Systematic and statistical uncertainties affecting the \Wgg and \Zgg
  fiducial cross section measurements, presented as percentages of the
  measured cross section.}
\newcolumntype{.}{D{.}{.}{2.1}}
\begin{tabular}{l....}
\hline \hline
 & \multicolumn{2}{c}{\Wgg} & \multicolumn{2}{c}{\Zgg}  \\
 &\multicolumn{1}{c}{$\Pe$ channel} & \multicolumn{1}{c}{$\mu$ channel} & \multicolumn{1}{c}{$\Pe\Pe$ channel} & \multicolumn{1}{c}{$\mu\mu$ channel}  \\
\hline
 \multicolumn{5}{c}{Systematic uncertainties associated with the simulation} \\  \hline
 Simulation statistical uncertainty & 2.8 & 2.4 & 3.3 & 2.9 \\
 Trigger & 0.5 & 0.3 & 1.3 & 1.2 \\
 Lepton and photon ID and energy scale & 4.1 & 3.0 & 5.3 & 4.3 \\
 \ptmiss scale & 1.5 & 1.4 &\multicolumn{1}{c}{ \NA} & \multicolumn{1}{c}{\NA} \\
 Pileup & 0.5 & 0.2 & 1.3 & 0.4 \\
 PDFs, renorm. and fact. scales & 1.5 & 1.6 & 1.2 & 1.3 \\   \hline
 \multicolumn{5}{c}{Systematic uncertainties associated with backgrounds} \\  \hline
 Misidentified jet & 36.6 & 37.2 & 15.1 & 12.5 \\
 Misidentified electron & 6.9 & \multicolumn{1}{c}{\NA}  & \multicolumn{1}{c}{\NA} & \multicolumn{1}{c}{\NA} \\
 Prompt diphoton & 6.7 & 5.8  & 0.2 & 0.3 \\
\hline\hline
 \multicolumn{5}{c}{Summary} \\  \hline
 Total statistical & 47.8 & 29.6  & 16.6 & 13.7 \\
 Total systematic & 38.3 & 37.9 & 16.5 & 13.7 \\
 Integrated luminosity & 2.6 & 2.6  & 2.6 & 2.6 \\
\hline \hline
\end{tabular}
\end{center}
\end{table}

The cross sections measured in the electron and muon channels of each
analysis are combined, assuming lepton universality, using the method
of best linear unbiased
estimates~\cite{LyonsBLUE,ValassiBLUE,NisiusBLUE}, thereby decreasing
the statistical uncertainties. We measure fiducial cross sections
of \wggxscomb and \zggxscomb for the \Wgg and \Zgg processes,
respectively. The measured cross sections are in agreement with the
NLO theoretical predictions of $4.8\pm 0.5\unit{fb}$ and $13.0\pm
1.5\unit{fb}$ for the \Wgg and \Zgg final states, respectively.  The
predicted cross sections are calculated within the fiducial phase
space given in Table~\ref{tab:fid_def} using \MGvNLO.
Table~\ref{tab:cross_sections} summarizes these results.

\begin{table}[tbph]
\begin{center}
\topcaption{\label{tab:cross_sections}
  Measured fiducial cross section for each channel and for the
  combination of channels for the \Wgg and \Zgg analyses. The combined
  cross sections assume lepton universality and are given for the
  decay to a single lepton family ($\ell$). The predictions are
  reported as well.}
\begingroup
\renewcommand*{\arraystretch}{1.2}
\begin{tabular}{lcc}
\hline\hline
Channel & Measured fiducial cross section \\
\hline
$\Wgg \to \Pe^{\pm}\nu\gamma\gamma$          & \wggxsel   \\
$\Wgg \to \mu^{\pm}\nu\gamma\gamma  $               & \wggxsmu   \\
$\Wgg \to \ell^{\pm}\nu\gamma\gamma $               & \wggxscomb \\
\hline
$\Zgg \to \Pep\Pem\gamma\gamma$ & \zggxsel   \\
$\Zgg \to \mu^{+}\mu^{-}\gamma\gamma  $             & \zggxsmu   \\
$\Zgg \to \ell^{+}\ell^{-}\gamma\gamma$             & \zggxscomb \\
\hline\hline
Channel & Prediction \\
\hline
$\Wgg \to \ell^{\pm}\nu\gamma\gamma $   &  $4.8\pm 0.5\fb$ \\
$\Zgg \to \ell^{+}\ell^{-}\gamma\gamma$ & $13.0\pm 1.5\fb$ \\
\hline
\hline
\end{tabular}
\endgroup
\end{center}
\end{table}

\section{Limits on aQGCs}

\begin{figure}[thbp]
\centering
\includegraphics[width=0.7\textwidth]{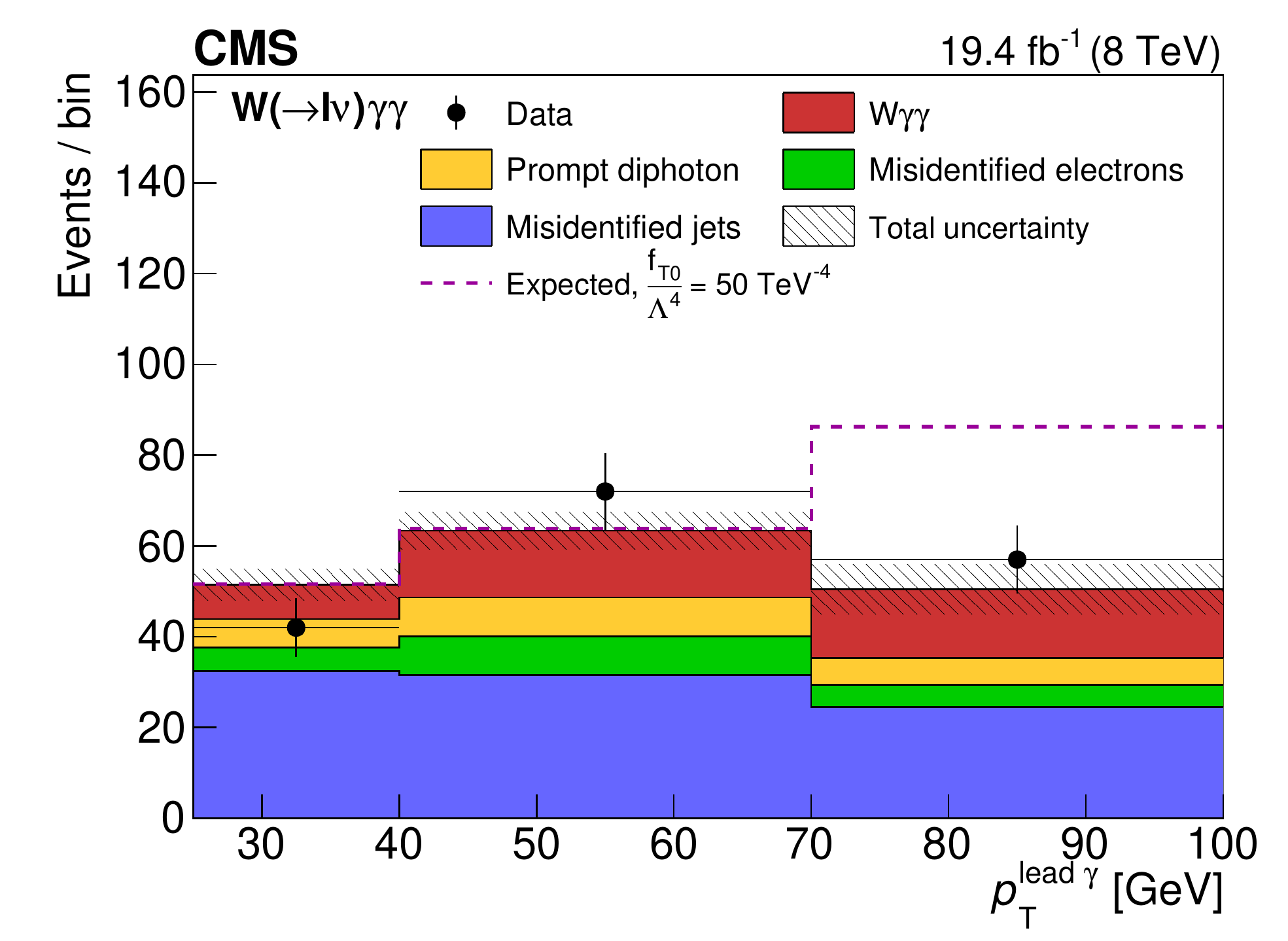}
\caption{\label{fig:final_plot_with_lt0_50}
  Distributions of the leading photon \pt for the \Wgg analysis with
  the electron and muon channels summed. The points display the
  observed data and the histograms give the predictions for the
  background and signal. The indicated uncertainties in the data
  points are calculated using Poisson statistics. The hatched area
  displays the total uncertainty in the sum of these predictions. The
  expected distribution with the inclusion of an aQGC with
  $\ft{0}/\Lambda^{4} = 50\TeV^{-4}$ is shown as the dashed line. The
  last bin includes all events in which the leading photon \pt exceeds
  70\GeV.}
\end{figure}

Anomalous QGCs are modeled using a dimension-8 effective field theory
parametrization~\cite{Degrande}. The effective field theory extends
the SM Lagrangian to terms of dimension larger than four. Each
additional dimension is suppressed by a power of the energy scale
$\Lambda$ at which the new phenomena appear. The terms in the extended
Lagrangian having odd-numbered dimensionality lead to baryon and
lepton number violation and are therefore not considered here. The
dimension-8 term is then the lowest-dimension term that produces
aQGCs. Fourteen dimension-8 operators contribute to the
$\W\W\gamma\gamma$ vertex~\cite{Belanger:1999aw,Eboli:2006wa}. We
focus our study on the couplings that contain products of electroweak
field strength tensors, in particular those that are constrained by
this analysis: $\fm{2}$, $\fm{3}$, $\ft{0}$, $\ft{1}$, and
$\ft{2}$~\cite{Baak:2013fwa}. Anomalous QGCs enhance the production of
signal events at high momentum scales. To increase sensitivity to
these enhancements, limits on aQGCs are obtained using only events in
which the leading-photon \pt exceeds
70\GeV. Figure~\ref{fig:final_plot_with_lt0_50} shows the predicted
yield from an aQGC with $\ft{0}/\Lambda^{4} = 50\TeV^{-4}$, compared
to the signal and background predictions for the sum of the electron
and muon channels. A profile likelihood is used to establish 95\%
confidence level (CL) intervals for the aQGC parameters. Each coupling
is profiled individually, with the other couplings set to their SM
values. Since all couplings predict an excess of the data at large photon
\pt, the observed limits are larger than the expected limits for all
couplings. The resulting limits are reported in Table~\ref{tab:aqgc}.

\begin{table}[thbp]
\begin{center}
\topcaption{\label{tab:aqgc}
  Expected and observed 95\% CL limits on anomalous quartic gauge
  couplings. Limits are obtained using \Wgg events in which the leading
  photon \pt exceeds 70\GeV.}
\begingroup
\renewcommand*{\arraystretch}{1.2}
\begin{tabular}{lcc}
\hline\hline
\Wgg & Expected ($\TeVns^{-4}$) & Observed ($\TeVns^{-4}$)\\
\hline
$\fm{2}/\Lambda^{4}$ & $[-549, 531  ]$ & $[-701, 683  ]$ \\
$\fm{3}/\Lambda^{4}$ & $[-916, 950  ]$ & $[-1170, 1220]$ \\
$\ft{0}/\Lambda^{4}$ & $[-26.5, 27.0]$ & $[-33.5, 34.0]$ \\
$\ft{1}/\Lambda^{4}$ & $[-34.5, 34.8]$ & $[-44.3, 44.8]$ \\
$\ft{2}/\Lambda^{4}$ & $[-74.6, 73.7]$ & $[-93.8, 93.2]$ \\
\hline
\hline
\end{tabular}
\endgroup
\end{center}
\end{table}

\section{Summary\label{sec:conclusions}}

Cross sections have been measured for \Wgg and \Zgg production
in \Pp\Pp\ collisions at $\sqrt{s}=8\TeV$ using data corresponding
to an integrated luminosity of 19.4\fbinv collected with the CMS
experiment. The cross sections were measured in fiducial regions that
are defined by criteria similar to those used to select signal
events. The fiducial cross sections are defined for \W and \Z boson
decays to a single lepton family. The measured fiducial cross
sections for these final states are, respectively, \wggxscombsum
and \zggxscombsum, consistent with the NLO theoretical predictions of
$4.8\pm 0.5\fb$ and $13.0\pm 1.5\fb$. These measurements correspond to
significances for observing the signal of \wggsignosig
and \zggsignosig standard deviations for the \Wgg and \Zgg final
states, respectively. 
In comparison, the ATLAS experiment measured the \Wgg and \Zgg final 
states with significances of greater than three standard deviations 
and equal to 6.3 standard deviations, respectively~\cite{Aad:2015uqa,Aad:2016sau}.  
The \Wgg final state is used to place limits at
95\% CL on anomalous quartic gauge couplings using a dimension-8
effective field theory. In particular, stringent limits are placed on
the $\ft{0}$ coupling parameter of \ftResult.

\begin{acknowledgments}
We congratulate our colleagues in the CERN accelerator departments for the excellent performance of the LHC and thank the technical and administrative staffs at CERN and at other CMS institutes for their contributions to the success of the CMS effort. In addition, we gratefully acknowledge the computing centers and personnel of the Worldwide LHC Computing Grid for delivering so effectively the computing infrastructure essential to our analyses. Finally, we acknowledge the enduring support for the construction and operation of the LHC and the CMS detector provided by the following funding agencies: BMWFW and FWF (Austria); FNRS and FWO (Belgium); CNPq, CAPES, FAPERJ, and FAPESP (Brazil); MES (Bulgaria); CERN; CAS, MoST, and NSFC (China); COLCIENCIAS (Colombia); MSES and CSF (Croatia); RPF (Cyprus); SENESCYT (Ecuador); MoER, ERC IUT, and ERDF (Estonia); Academy of Finland, MEC, and HIP (Finland); CEA and CNRS/IN2P3 (France); BMBF, DFG, and HGF (Germany); GSRT (Greece); OTKA and NIH (Hungary); DAE and DST (India); IPM (Iran); SFI (Ireland); INFN (Italy); MSIP and NRF (Republic of Korea); LAS (Lithuania); MOE and UM (Malaysia); BUAP, CINVESTAV, CONACYT, LNS, SEP, and UASLP-FAI (Mexico); MBIE (New Zealand); PAEC (Pakistan); MSHE and NSC (Poland); FCT (Portugal); JINR (Dubna); MON, RosAtom, RAS, RFBR and RAEP (Russia); MESTD (Serbia); SEIDI, CPAN, PCTI and FEDER (Spain); Swiss Funding Agencies (Switzerland); MST (Taipei); ThEPCenter, IPST, STAR, and NSTDA (Thailand); TUBITAK and TAEK (Turkey); NASU and SFFR (Ukraine); STFC (United Kingdom); DOE and NSF (USA).

\hyphenation{Rachada-pisek} Individuals have received support from the Marie-Curie program and the European Research Council and EPLANET (European Union); the Leventis Foundation; the A. P. Sloan Foundation; the Alexander von Humboldt Foundation; the Belgian Federal Science Policy Office; the Fonds pour la Formation \`a la Recherche dans l'Industrie et dans l'Agriculture (FRIA-Belgium); the Agentschap voor Innovatie door Wetenschap en Technologie (IWT-Belgium); the Ministry of Education, Youth and Sports (MEYS) of the Czech Republic; the Council of Science and Industrial Research, India; the HOMING PLUS program of the Foundation for Polish Science, cofinanced from European Union, Regional Development Fund, the Mobility Plus program of the Ministry of Science and Higher Education, the National Science Center (Poland), contracts Harmonia 2014/14/M/ST2/00428, Opus 2014/13/B/ST2/02543, 2014/15/B/ST2/03998, and 2015/19/B/ST2/02861, Sonata-bis 2012/07/E/ST2/01406; the National Priorities Research Program by Qatar National Research Fund; the Programa Clar\'in-COFUND del Principado de Asturias; the Thalis and Aristeia programs cofinanced by EU-ESF and the Greek NSRF; the Rachadapisek Sompot Fund for Postdoctoral Fellowship, Chulalongkorn University and the Chulalongkorn Academic into Its 2nd Century Project Advancement Project (Thailand); and the Welch Foundation, contract C-1845.
\end{acknowledgments}

\bibliography{auto_generated}

\cleardoublepage \appendix\section{The CMS Collaboration \label{app:collab}}\begin{sloppypar}\hyphenpenalty=5000\widowpenalty=500\clubpenalty=5000\input{SMP-15-008-authorlist.tex}\end{sloppypar}
\end{document}

%% file: SMP-15-008-authorlist.tex
\textbf{Yerevan Physics Institute,  Yerevan,  Armenia}\\*[0pt]
A.M.~Sirunyan, A.~Tumasyan
\vskip\cmsinstskip
\textbf{Institut f\"{u}r Hochenergiephysik,  Wien,  Austria}\\*[0pt]
W.~Adam, E.~Asilar, T.~Bergauer, J.~Brandstetter, E.~Brondolin, M.~Dragicevic, J.~Er\"{o}, M.~Flechl, M.~Friedl, R.~Fr\"{u}hwirth\cmsAuthorMark{1}, V.M.~Ghete, C.~Hartl, N.~H\"{o}rmann, J.~Hrubec, M.~Jeitler\cmsAuthorMark{1}, A.~K\"{o}nig, I.~Kr\"{a}tschmer, D.~Liko, T.~Matsushita, I.~Mikulec, D.~Rabady, N.~Rad, B.~Rahbaran, H.~Rohringer, J.~Schieck\cmsAuthorMark{1}, J.~Strauss, W.~Waltenberger, C.-E.~Wulz\cmsAuthorMark{1}
% ...
\vskip\cmsinstskip
% ...
1:~~Also at Vienna University of Technology, Vienna, Austria\\
% ...
